\documentstyle[12pt]{article}

\font\twlgot =eufm10 scaled \magstep1 \font\egtgot =eufm8
\font\sevgot =eufm7 \font\twlmsb =msbm10 scaled \magstep1
\font\egtmsb =msbm8 \font\sevmsb =msbm7

\newfam\gotfam

\textfont\gotfam\twlgot \scriptfont\gotfam\egtgot
\scriptscriptfont\gotfam\sevgot

\newfam\msbfam
\textfont\msbfam\twlmsb \scriptfont\msbfam\egtmsb
\scriptscriptfont\msbfam\sevmsb
\def\Bbb{\protect\pBbb}
\def\pBbb{\relax\ifmmode\expandafter\Bb\else\typeout{You cann't use
Bbb in text mode}\fi}
\def\Bb #1{{\fam\msbfam\relax#1}}

\def\thebibliography#1{\bigskip\section*{}\bigskip\list
{$^{\arabic{enumi}}$}{\settowidth\labelwidth{#1}\leftmargin\labelwidth
\advance\leftmargin\labelsep
\usecounter{enumi}}
\def\newblock{\hskip .11em plus .33em minus .07em}
\sloppy\clubpenalty4000\widowpenalty4000 \sfcode`\.=1000\relax}

\def\op#1{\mathop{\fam0 #1}\limits}

\newcommand{\id}{{\rm Id\,}}

\newcommand{\beq}{\begin{equation}}
\newcommand{\eeq}{\end{equation}}
\newcommand{\ben}{\begin{eqnarray}}
\newcommand{\een}{\end{eqnarray}}
\newcommand{\be}{\begin{eqnarray*}}
\newcommand{\ee}{\end{eqnarray*}}
\newcommand{\bea}{\begin{eqalph}}
\newcommand{\eea}{\end{eqalph}}

\newcommand{\cV}{{\cal V}}

\newcommand{\cH}{{\cal H}}

\newcommand{\cF}{{\cal F}}

\newcommand{\bL}{{\bf L}}

\newcommand{\al}{\alpha}

\newcommand{\bt}{\beta}

\newcommand{\la}{\lambda}

\newcommand{\f}{\phi}

\newcommand{\Om}{\Omega}

\newcommand{\g}{\gamma}

\newcommand{\F}{\Phi}

\newcommand{\vt}{\vartheta}

\newcommand{\di}{{\rm dim\,}}

\newcommand{\w}{\wedge}
\newcommand{\wt}{\widetilde}

\newcommand{\ol}{\overline}
\newcommand{\dr}{\partial}
\newcommand{\ar}{\op\longrightarrow}

\let\ssection=\section
\renewcommand{\section}{\setcounter{equation}{0}\ssection}

\newcounter{eqalph}
\newcounter{equationa}
\newcounter{remark}
\newcounter{example}
\newcounter{theorem}
\newcounter{proposition}
\newcounter{lemma}
\newcounter{corollary}
\newcounter{definition}
\newenvironment{eqalph}{\stepcounter{equation}
\setcounter{equationa}{\value{equation}} \setcounter{equation}{0}

\begin{eqnarray}}{\end{eqnarray}\setcounter{equation}{\value{equationa}}}

\def\theremark{\arabic{remark}}

\def\thetheorem{\arabic{theorem}}

\newenvironment{theo}{\refstepcounter{theorem}
{\bf Theorem \thetheorem:}}{}

\newenvironment{defi}{\refstepcounter{theorem}
{\bf Definition \thetheorem:}}{}

\newcommand{\mar}[1]{}

\begin{document}
\hbox{}

{\parindent=0pt

{\large\bf Superintegrable non-autonomous Hamiltonian systems}
\bigskip

{\sc G. Sardanashvily}

{\sl Department of Theoretical Physics, Moscow State University,
117234 Moscow, Russia}

\bigskip
\bigskip

The Mishenko--Fomenko theorem on action-angle coordinates for
superintegrable autonomous Hamiltonian systems is extended to the
non-autonomous ones.
 }

\bigskip
\bigskip

\noindent {\bf I. INTRODUCTION}
\bigskip

The Liouville -- Arnold theorem for completely integrable
systems$^{1-3}$ and the Mishchenko -- Fomenko theorem for the
superintegrable ones$^{4-6}$ state the existence of action-angle
coordinates around a compact invariant submanifold. These theorems
were generalized to the case of non-compact invariant
submanifolds.$^{7-10}$ This generalization enable us to analyze
completely integrable non-autonomous Hamiltonian systems whose
invariant submanifolds are necessarily non-compact.$^{11,12}$ Here
we aim to extend this analysis to superintegrable non-autonomous
Hamiltonian systems.

We consider a non-autonomous mechanical system whose configuration
space is a fibre bundle $Q\to \Bbb R$ over the time axis $\Bbb R$
endowed with the Cartesian coordinate $t$ possessing transition
functions $t'=t+$const. Its phase space is the vertical cotangent
bundle $V^*Q\to Q$ of $Q\to\Bbb R$ endowed with the Poisson
structure $\{,\}_V$ (\ref{m72}).$^{13,14}$ A Hamiltonian of a
non-autonomous mechanical system is a section $H$ (\ref{ws513}) of
the one-dimensional fibre bundle
\mar{z11'}\beq
\zeta:T^*Q\to V^*Q, \label{z11'}
\eeq
where $T^*Q$ is the cotangent bundle of $Q$ endowed with the
canonical symplectic form (\ref{09130}).

\begin{defi} \label{i0} \mar{i0}
A non-autonomous Hamiltonian  system of $m=\di Q-1$ degrees of
freedom is called superintegrable if admits $m\leq n<2m$ integrals
of motion $\F_1,\ldots,\F_n$ obeying the following conditions.

(i) All the functions $\F_\al$ are independent, i.e. the $n$-form
$d\F_1\w\cdots\w d\F_n$ nowhere vanishes on $V^*Q$. It follows
that the map
\mar{nc4}\beq
\F:V^*Q\to N=(\F_1(V^*Q),\ldots,\F_n(V^*Q))\subset \Bbb R^n
\label{nc4}
\eeq
is a fibred manifold over a connected open subset $N\subset\Bbb
R^n$.

(ii) There exist smooth real functions $s_{ij}$ on $N$ such that
\mar{nc1}\beq
\{\F_\al,\F_\bt\}_V= s_{\al\bt}\circ \F, \qquad \al,\bt=1,\ldots,
n. \label{nc1}
\eeq

(iii) The matrix function with the entries $s_{\al\bt}$
(\ref{nc1}) is of constant corank $k=2m-n$ at all points of $N$.
\end{defi}

To describe this non-autonomous Hamiltonian system, we use the
fact that there exists an equivalent autonomous Hamiltonian system
on the cotangent bundle $T^*Q$ (Theorem \ref{09121}) which is
superintegrable (Theorem \ref{09141}). Our goal is the following.

\begin{theo} \label{nc0'} \mar{nc0'}
Let Hamiltonian vector fields of the functions $\F_\al$ be
complete, and let fibres of the fibred manifold $\F$ (\ref{nc4})
be connected and mutually diffeomorphic. Then there exists an open
neighborhood $U_M$ of a fibre $M$ of $\F$ (\ref{nc4}) which is a
trivial principal bundle with the structure group
\mar{g120}\beq
\Bbb R^{1+k-r}\times T^r \label{g120}
\eeq
whose bundle coordinates are the generalized action-angle
coordinates
\mar{09135}\beq
(p_A,q^A,I_\la,t,y^\la), \qquad A=1,\ldots,n-m, \qquad
\la=1,\ldots, k,\label{09135}
\eeq
such that:

(i) $(t,y^\la)$ are coordinates on the toroidal cylinder
(\ref{g120}),

(ii) the Poisson bracket $\{,\}_V$ on $U_M$ reads
\be
\{f,g\}_V = \dr^Af\dr_Ag-\dr^Ag\dr_Af + \dr^\la f\dr_\la g-\dr^\la
g\dr_\la f,
\ee

(iii) the Hamiltonian $H$ depends only on the action coordinates
$I_\la$,

(iv) the integrals of motion $\F_1, \ldots \F_n$ are independent
of coordinates $(t,y^\la)$.
\end{theo}

If $n=m$, we are in the case of a completely integrable
non-autonomous Hamiltonian system (Theorem \ref{z13}).

\bigskip
\bigskip

\noindent {\bf II. NON-AUTONOMOUS HAMILTONIAN MECHANICS}
\bigskip

The configuration space $Q\to \Bbb R$ of a non-autonomous
Hamiltonian system is equipped with bundle coordinates $(t,q^i)$,
$i=1,\ldots,m$.$^{13,14}$ Its phase space $V^*Q$ is provided with
holonomic coordinates $(t, q^i,p_i=\dot q_i)$ with respect to
fibre bases $\{\ol dq^i\}$ for $V^*Q$. The cotangent bundle $T^*Q$
of $Q$ plays a role of the homogeneous phase space endowed with
holonomic coordinates $(t,q^i,p_0,p_i)$ possessing transition
functions
\mar{z40}\beq
p'_i = \frac{\dr q^j}{\dr{q'}^i}p_j, \qquad p'_0=p_0+ \frac{\dr
q^j}{\dr t'}p_j. \label{z40}
\eeq
The cotangent bundle $T^*Q$ admits the canonical symplectic form
\mar{09130}\beq
\Om=dp_0\w dt +dp_i\w dq^i, \label{09130}
\eeq
and the corresponding Poisson bracket
\be
\{f,g\} =\dr^0f\dr_tg - \dr^0g\dr_tf +\dr^if\dr_ig-\dr^ig\dr_if,
\qquad f,g\in C^\infty(T^*Q).
\ee

A glance at the transformation law (\ref{z40}) shows that the
one-dimensional fibre bundle $\zeta$ (\ref{z11'}) is a trivial
affine bundle. Given its global section
\mar{ws513}\beq
H:V^*Q\to T^*Q, \qquad p_0\circ H=-\cH(t,q^j,p_j), \label{ws513}
\eeq
the cotangent bundle $T^*Q$ is equipped with the fibre coordinate
\mar{09151}\beq
I_0=p_0+\cH, \qquad I_0\circ H=0, \label{09151}
\eeq
possessing the identity transition functions. With respect to the
coordinates
\mar{09150}\beq
(t,q^i,I_0,p_i), \qquad i=1,\ldots,m, \label{09150}
\eeq
the fibration (\ref{z11'}) reads
\mar{z11}\beq
\zeta: \Bbb R\times V^*Q \ni (t,q^i,I_0,p_i)\to (t,q^i,p_i)\in
V^*Q. \label{z11}
\eeq

The fibre bundle (\ref{z11'}) provides the vertical cotangent
bundle $V^*Q$ with the canonical Poisson structure $\{,\}_V$ such
that
\mar{m72',72}\ben
&& \zeta^*\{f,g\}_V=\{\zeta^*f,\zeta^*g\}, \label{m72'}\\
&& \{f,g\}_V = \dr^if\dr_ig-\dr^ig\dr_if, \qquad f,g\in
C^\infty(V^*Q). \label{m72}
\een
The Hamiltonian vector fields of functions on $V^*Q$ with respect
to the Poisson bracket (\ref{m72}) are vertical vector fields
\mar{093,4}\ben
&& \vt_f = \dr^if\dr_i- \dr_if\dr^i, \qquad f\in C^\infty(V^*Q),
\label{093} \\
&& [\vt_f,\vt_{f'}]=\vt_{\{f,f'\}_V}, \label{094}
\een
on $V^*Q\to \Bbb R$. Accordingly, the corresponding symplectic
foliation on the phase space $V^*Q$ coincides with the fibration
$V^*Q\to \Bbb R$.

A Hamiltonian of non-autonomous mechanics on the phase space
$V^*Q$ is defined as a global section (\ref{ws513}) of the affine
bundle $\zeta$ (\ref{z11'}). Then there exists a unique vector
field $\g_H$ on $V^*Q$ such that
\mar{z3}\ben
&&\g_H\rfloor dt=1, \qquad \g_H\rfloor H^*\Om=0, \nonumber\\
&&\g_H=\dr_t + \dr^i\cH\dr_i- \dr_i\cH\dr^i. \label{z3}
\een
This vector field, called the Hamilton vector field, defines the
first order Hamilton equations
\mar{zz20}\beq
q^i_t=\dr^i\cH, \qquad p_{ti}=-\dr_i\cH \label{zz20}
\eeq
on a fibre bundle $V^*Q\to\Bbb R$ with respect to the adapted
coordinates $(t,q^i,p_i,q^i_t,p_{ti})$ on the first order jet
manifold $J^1V^*Q$ of $V^*Q\to\Bbb R$. Due to the canonical
imbedding $J^1V^*Q\to TV^*Q$, the Hamilton equations (\ref{zz20})
are equivalent to the first order differential equations
\mar{z20}\beq
\dot t=1, \qquad \dot q^i=\dr^i\cH, \qquad \dot p_i=-\dr_i\cH
\label{z20}
\eeq
on a manifold $V^*Q$.

In order to describe evolution of a mechanical system, the
Hamilton vector field $\g_H$ (\ref{z3}) is assumed to be complete.
It defines a trivialization
\be
V^*Q\cong\Bbb R\times P
\ee
which is a canonical automorphism of the Poisson manifold $V^*Q$
such that the corresponding coordinates $(t,\ol q^i,\ol p_i)$ are
the initial date coordinates.$^{14}$ With respect to these
coordinates, the Hamiltonian (\ref{ws513}) reads $\cH=0$, and the
Hamilton equations (\ref{zz20}) take the form
\be
\ol q^i_t=0, \qquad \ol p_{ti}=0.
\ee

We agree to call $(V^*Q,H)$ the non-autonomous Hamiltonian system
of $m$ degrees of freedom.

\begin{theo} \label{09121} \mar{09121} A non-autonomous Hamiltonian system $(V^*Q,H)$
is equivalent to an autonomous Hamiltonian system $(T^*Q,\cH^*)$
of $m+1$ degrees of freedom on a symplectic manifold $(T^*Q,\Om)$
whose Hamiltonian is the function$^{15,16}$
\mar{mm16}\beq
\cH^*=I_0=p_0+\cH. \label{mm16}
\eeq
\end{theo}

The Hamiltonian vector field $u_{\cH^*}$ of $\cH^*$ (\ref{mm16})
on $T^*Q$ is
\mar{z5}\beq
u_{\cH^*}=\dr_t -\dr_t\cH\dr^0+ \dr^i\cH\dr_i- \dr_i\cH\dr^i.
\label{z5}
\eeq
Written relative to the coordinates (\ref{09150}), this vector
field reads
\mar{z5'}\beq
u_{\cH^*}=\dr_t + \dr^i\cH\dr_i- \dr_i\cH\dr^i. \label{z5'}
\eeq
It is projected onto the Hamilton vector field $\g_H$ (\ref{z3})
on $V^*Q$ such that
\mar{ws525}\beq
\zeta^*(\bL_{\g_H}f)=\{\cH^*,\zeta^*f\}, \qquad f\in
C^\infty(V^*Q). \label{ws525}
\eeq
The corresponding autonomous Hamilton equations on $T^*Q$ take the
form
\mar{z20'}\beq
\dot t=1, \qquad \dot p_0=-\dr_t\cH, \qquad \dot q^i=\dr^i\cH,
\qquad \dot p_i=-\dr_i\cH. \label{z20'}
\eeq
They are equivalent to the Hamilton equations (\ref{z20}).

Obviously, the vector field $u_{\cH^*}$ (\ref{z5'}) is complete if
the Hamilton vector field $\g_H$ (\ref{z3}) is complete.

\bigskip
\bigskip

\noindent {\bf III. COMPLETELY INTEGRABLE NON-AUTONOMOUS
HAMILTONIAN SYSTEMS}
\bigskip

An integral of motion of a non-autonomous Hamiltonian system
$(V^*Q,H)$ is defined as a smooth real function $F$ on $V^*Q$
whose Lie derivative
\be
\bL_{\g_H} F=\g_H\rfloor dF=\dr_tF +\{\cH,F\}_V
\ee
along the Hamilton vector field $\g_H$ (\ref{z3}) vanishes. Given
the Hamiltonian vector field $\vt_F$ (\ref{093}) of $F$ with
respect to the Poisson bracket (\ref{m72}), it is easily justified
that
\mar{092}\beq
[\g_H,\vt_F]=\vt_{\bL_{\g_H} F}. \label{092}
\eeq

\begin{defi} \label{09122} \mar{09122}
A non-autonomous Hamiltonian system $(V^*Q,H)$ of $m$ degrees of
freedom is said to be completely integrable if it admits $m$
integrals of motion $F_1,\ldots,F_m$ which are in involution with
respect to the Poisson bracket $\{,\}_V$ (\ref{m72}) and whose
differentials $dF_\al$ are linearly independent, i.e.
\be
dF_1\w\cdots\w dF_m\neq 0.
\ee
\end{defi}

By virtue of the relations (\ref{094}) and (\ref{092}), the vector
fields
\mar{095}\ben
&& (\g_H,\vt_{F_1},\ldots,\vt_{F_m}), \label{095}\\
&& \vt_{F_\al} = \dr^iF_\al\dr_i- \dr_iF_\al\dr^i. \nonumber
\een
mutually commute and, therefore, they span an $(m+1)$-dimensional
involutive distribution $\cV$ on $V^*Q$. Let $G$ be the group of
local diffeomorphisms of $V^*Q$ generated by the flows of vector
fields (\ref{095}). Maximal integral manifolds of $\cV$ are the
orbits of $G$ and invariant submanifolds of vector fields
(\ref{095}).$^{12,17}$ They yield a foliation $\cF$ of $V^*Q$.

Let $(V^*Q,H)$ be a non-autonomous Hamiltonian system and
$(T^*Q,\cH^*)$ an equivalent autonomous Hamiltonian system on
$T^*Q$. An immediate consequence of the relations (\ref{m72'}) and
(\ref{ws525}) is the following.$^9$

\begin{theo} \label{z6} \mar{z6}
Given a completely integrable non-autonomous Hamiltonian system
\mar{097'}\beq
(\g_H,F_1,\ldots,F_m) \label{097'}
\eeq
of $m$ degrees of freedom on $V^*Q$, the autonomous Hamiltonian
system
\mar{097}\beq
(\cH^*,\zeta^*F_1,\ldots,\zeta^*F_m) \label{097}
\eeq
of $m+1$ degrees of freedom on $T^*Q$ is completely integrable.
\end{theo}

The Hamiltonian vector fields
\mar{099}\ben
&& (u_{\cH^*},u_{\zeta^*F_1},\ldots,u_{\zeta^*F_m}), \label{099}\\
&& u_{\zeta^*F_\al} = \dr^iF_\al\dr_i- \dr_iF_\al\dr^i, \nonumber
\een
of the autonomous integrals of motion (\ref{097}) span an
$(m+1)$-dimensional involutive distribution $\cV_T$ on $T^*Q$ such
that
\mar{098}\beq
T\zeta(\cV_T)=\cV, \qquad TH(\cV)=\cV_T|_{H(V^*Q)=I_0=0},
\label{098}
\eeq
where
\be
TH: TV^*Q\ni(t,q^i,p_i,\dot t,\dot q^i,\dot
p_i)\to(t,q^i,p_i,I_0=0,\dot t,\dot q^i,\dot p_i,\dot I_0=0)\in
TT^*Q.
\ee
It follows that, if $M$ is an invariant submanifold of the
completely integrable non-autonomous Hamiltonian system
(\ref{097'}), then $H(M)$ is an invariant submanifold of the
completely integrable autonomous Hamiltonian system (\ref{097}).

In order do introduce generalized action-angle coordinates around
an invariant submanifold $M$ of the completely integrable
non-autonomous Hamiltonian system (\ref{097'}), let us suppose
that the vector fields (\ref{095}) on $M$ are complete. It follows
that $M$ is a locally affine manifold diffeomorphic to a toroidal
cylinder
\mar{0111}\beq
(\Bbb R^{1+m-r}\times T^r). \label{0111}
\eeq
Moreover, let assume that there exists an open neighbourhood $U$
of $M$ such that the foliation $\cF$ of $U$ is a fibred manifold
$\f: U\to N$ over a domain $N\subset \Bbb R^m$ whose fibres are
mutually diffeomorphic.$^{11}$

Because the morphism $TH$ (\ref{098}) is a bundle isomorphism, the
Hamiltonian vector fields (\ref{099}) on the invariant submanifold
$H(M)$ of the completely integrable autonomous Hamiltonian system
are complete. Since the affine bundle $\zeta$ (\ref{z11}) is
trivial, the open neighbourhood $\zeta^{-1}(U)$ of the invariant
submanifold $H(M)$ is a fibred manifold
\be
\wt\f: \zeta^{-1}(U)= \Bbb R\times U \ar^{(\id\Bbb R,\f)} \Bbb
R\times N = N'
\ee
over a domain $N'\subset \Bbb R^{m+1}$ whose fibres are
diffeomorphic to the toroidal cylinder (\ref{0111}). In accordance
with the Liouville -- Arnold theorem extended to the case of
non-compact invariant submanifolds,$^7$ the open neighbourhood
$\zeta^{-1}(U)$ of $H(M)$ is a trivial principal bundle
\mar{0910}\beq
\zeta^{-1}(U)=N'\times (\Bbb R^{1+m-r}\times T^r)\to N'
\label{0910}
\eeq
with the structure group (\ref{0111}) whose bundle coordinates are
the generalized action-angle coordinates
\mar{0911}\beq
(I_0,I_1,\ldots,I_m, t,z^1,\ldots,z^m) \label{0911}
\eeq
such that:

(i) $(t,z^a)$ are coordinates on the toroidal cylinder
(\ref{0111}),

(ii) the symplectic form (\ref{09130}) on $\zeta^{-1}(U)$ reads
\be
\Om=dI_0\w dt + dI_a\w dz^a,
\ee

(iii) $\cH^*=I_0$,

(iv) the integrals of motion $\zeta^*F_1,\ldots,\zeta^*F_m$ depend
only on the action coordinates $I_1,\ldots,I_m$.

Provided with the coordinates (\ref{0911}), $\zeta^{-1}(U)=
U\times\Bbb R$ is a trivial bundle possessing the fibre coordinate
$I_0$ (\ref{09151}). Consequently, the open neighbourhood $U$ of
an invariant submanifold $M$ of the completely integrable
non-autonomous Hamiltonian system (\ref{095}) is diffeomorphic to
the Poisson annulus
\mar{0915}\beq
U=N\times (\Bbb R^{1+m-r}\times T^r) \label{0915}
\eeq
endowed with the generalized action-angle coordinates
\mar{0916}\beq
(I_1,\ldots,I_m, t,z^1,\ldots,z^m) \label{0916}
\eeq
such that:

(i) the Poisson structure (\ref{m72}) on $U$ takes the form
\be
\{f,g\}_V = \dr^af\dr_ag-\dr^ag\dr_af,
\ee

(ii) the Hamiltonian (\ref{ws513}) reads $\cH=0$,

(iii) the integrals of motion $F_1,\ldots,F_m$ depend only on the
action coordinates $I_1,\ldots,I_m$.

The Hamilton equations (\ref{zz20}) relative to the generalized
action-angle coordinates (\ref{0916}) take the form
\be
 z^a_t=0, \qquad I_{ta}=0.
\ee
It follows that the generalized action-angle coordinates
(\ref{0916}) are the initial date coordinates.

Note that the generalized action-angle coordinates (\ref{0916}) by
no means are unique. Given a smooth function $\cH'$ on $\Bbb R^m$,
one can provide $\zeta^{-1}(U)$ with the generalized action-angle
coordinates
\be
t, \qquad z'^a=z^a- t\dr^a\cH', \qquad I'_0=I_0+\cH'(I_b), \qquad
I'_a=I_a.
\ee
With respect to these coordinates, a Hamiltonian of the autonomous
Hamiltonian system on $\zeta^{-1}(U)$ reads $\cH'^*=I'_0-\cH'$. A
Hamiltonian of the non-autonomous Hamiltonian system on $U$
endowed with the generalized action-angle coordinates $(t,
z'^a,I_a)$ is $\cH'$.

Thus, the following has been proved.

\begin{theo} \label{z13} \mar{z13}
Let $(\g_H,F_1,\ldots,F_m)$ be a completely integrable
non-autonomous Hamiltonian system. Let $M$ be its invariant
submanifold such that the vector fields (\ref{095}) on $M$ are
complete and there exists an open neighbourhood $U$ of $M$ which
is a fibred manifold in mutually diffeomorphic invariant
submanifolds. Then $U$ is diffeomorphic to the Poisson annulus
(\ref{0915}), and it can be provided with the generalized
action-angle coordinates (\ref{0916}) such that the integrals of
motion $(F_1,\ldots,F_m)$ and the Hamiltonian $H$ depend only on
the action coordinates $I_1,\ldots,I_m$.
\end{theo}

\bigskip
\bigskip

\noindent {\bf IV. SUPERINTEGRABLE NON-AUTONOMOUS HAMILTONIAN
SYSTEMS}
\bigskip

Let $(\g_H,\F_1,\ldots,\F_n)$ be a superintegrable non-autonomous
Hamiltonian system in accordance with Definition \ref{i0}. The
associated autonomous Hamiltonian system on $T^*Q$ possesses $n+1$
integrals of motion
\mar{09136}\beq
(\cH^*,\zeta^*\F_1,\ldots,\zeta^*\F_n) \label{09136}
\eeq
with the following properties.

(i) The functions (\ref{09136}) are mutually independent, and the
map
\mar{09140}\ben
&& \wt\F:T^*Q\to
(\cH^*(T^*Q),\zeta^*\F_1(T^*Q),\ldots,\zeta^*\F_n(T^*Q))= \label{09140}\\
&& \qquad (I_0,\F_1(V^*Q),\ldots,\F_n(V^*Q))= \Bbb R\times
N=N'\nonumber
\een
is a fibred manifold.

(ii) The functions (\ref{09136}) obey the relations
\be
\{\zeta^*\F_\al,\zeta^*\F_\bt\}= s_{\al\bt}\circ \zeta^*\F,\qquad
\{\cH^*,\zeta^*\F_\al\}=s_{0\al}=0
\ee
so that the matrix function with the entries
$(s_{0\al},s_{\al\bt})$ on $N'$ is of constant corank $2m+1-n$.

Refereing to the definition of an autonomous superintegrable
system,$^{4-6}$ we come to the following.

\begin{theo} \label{09141} \mar{09141} Given a superintegrable
non-autonomous Hamiltonian system $(\g_H,\F_\al)$ on $V^*Q$, the
associated autonomous Hamiltonian system (\ref{09136}) on $T^*Q$
is superintegrable.
\end{theo}

There is the commutative diagram
\be
\begin{array}{rcccl}
&T^*Q &\ar^\zeta & V^*Q&\\
_{\wt \F}& \put(0,10){\vector(0,-1){20}} & & \put(0,10){\vector(0,-1){20}}&_\F\\
& N' &\ar^\xi & N&
\end{array}
\ee
where $\zeta$ (\ref{z11}) and
\be
\xi:N'=\Bbb R\times N\to N
\ee
are trivial bundles. It follows that the fibred manifold
(\ref{09140}) is the pull-back $\wt \F=\xi^* \F$ of the fibred
manifold $\F$ (\ref{nc4}) onto $N'$.

Let the conditions of Theorem \ref{nc0'} hold. If the Hamiltonian
vector fields
\be
(\g_H,\vt_{\F_1},\ldots,\vt_{\F_n}),\qquad, \vt_{\F_\al}=
\dr^i\F_\al\dr_i- \dr_i\F_\al\dr^i,
\ee
of integrals of motion $\F_\al$ on $V^*Q$ are complete, the
Hamiltonian vector fields
\be
(u_{\cH^*},u_{\zeta^*\F_1},\ldots,u_{\zeta^*\F_n}), \qquad
u_{\zeta^*\F_\al} = \dr^i\F_\al\dr_i- \dr_i\F_\al\dr^i,
\ee
on $T^*Q$ are complete. If fibres of the fibred manifold $\F$
(\ref{nc4}) are connected and mutually diffeomorphic, the fibres
of the fibred manifold $\wt\F$ (\ref{09140}) also are well.

Let $M$ be a fibre of $\F$ (\ref{nc4}) and $H(M)$ the
corresponding fibre of $\wt\F$ (\ref{09140}). In accordance with
the Mishchenko -- Fomenko theorem extended to the case of
non-compact invariant submanifolds,$^{9,10}$ there exists an open
neighbourhood $U'$ of $H(M)$ which is a trivial principal bundle
with the structure group (\ref{g120}) whose bundle coordinates are
the generalized action-angle coordinates
\mar{09135'}\beq
(p_A,q^A,I_0,I_\la,t,y^\la,), \qquad A=1,\ldots,n-m, \qquad
\la=1,\ldots, k,\label{09135'}
\eeq
such that:

(i) $(t,y^\la)$ are coordinates on the toroidal cylinder
(\ref{g120}),

(ii) the symplectic form $\Om$ (\ref{09130}) on $U'$ reads
\be
\Om= dp_A\w dq^A + dI_0\w dt + dI_\al\w dy^\al,
\ee

(iii) the action coordinates $(I_0,I_\al)$ are expressed into the
values of the Casimir functions $C_0=I_0$, $C_\al$ of the
coinduced Poisson structure
\be
w=\dr^A\w\dr_A
\ee
on $N'$,

(iv) the Hamiltonian $\cH^*$ depends on the action coordinates,
namely, $\cH^*=I_0$,

(iv) the integrals of motion $\zeta^*\F_1, \ldots \zeta^*\F_n$ are
independent of the coordinates $(t,y^\la)$.

Provided with the generalized action-angle coordinates
(\ref{09135'}), the above mentioned neighbourhood $U'$ is a
trivial bundle $U'=\Bbb R\times U_M$ where $U_M=\zeta(U')$ is an
open neighbourhood of the fibre $M$ of the fibre bundle $\F$
(\ref{nc4}). As a result, we come to Theorem \ref{nc0'}.


\begin{thebibliography}{ddd}

%1
\bibitem{arn} V.Arnold and A.Avez, {\it Ergodic Problems in Classical Mechanics}
(Benjamin, New York, 1968).

%2
\bibitem{arn1} V.Arnold (Ed.), {\it Dynamical Systems III, IV}
(Springer-Verlag, Berlin, 1990).

%3
\bibitem{laz} V.Lazutkin, {\it KAM Theory and Semiclassical
Approximations to Eigenfunctions} (Springer-Verlag, Berlin, 1993).

%4
\bibitem{mishc} A.Mishchenko  and A.Fomenko, Funct. Anal.
Appl. {\bf 12}, 113 (1978).

%5
\bibitem{bols03} A.Bolsinov and B.Jovanovi\'c, Ann. Global
Anal. Geom. {\bf 23}, 305 (2003).

%6
\bibitem{fasso05} F.Fass\'o, Acta Appl. Math. {\bf 87}, 93
(2005).

%7
\bibitem{fior} E.Fiorani, G.Giachetta and G.Sardanashvily, J. Phys. A
{\bf 36}, L101 (2003); {\it E-print arXiv:} math.DS/0210346.

%8
\bibitem{jmp03} G.Giachetta, L.Mangiarotti and G.Sardanashvily, J. Math. Phys.
{\bf 44}, 1984 (2003); {\it E-print arXiv} math.DS/0211463.

%9
\bibitem{fior2} E.Fiorani and G.Sardanashvily, J. Phys. A
{\bf 39}, 14035 (2006); {\it E-print arXiv:} math.DG/0604104.

%10
\bibitem{jmp07} E.Fiorani and G.Sardanashvily, J. Math. Phys. {\bf
48}, 032001 (2007); {\it E-print arXiv:} math.DG/0610790.

%11
\bibitem{jpa02} G.Giachetta, L.Mangiarotti and G.Sardanashvily, J. Phys. A
 {\bf 35}, L439 (2002); {\it
E-print arXiv:} math.DS/0204151.

%12
\bibitem{book05} G.Giachetta, L.Mangiarotti and G.Sardanashvily, {\it
Geometric and Algebraic Topological Methods in Quantum Mechanics}
(World Scientific, Singapore, 2005).

%13
\bibitem{jmp98} G.Sardanashvily, J. Math. Phys. {\bf
39}, 2714 (1998).

%14
\bibitem{book98} L.Mangiarotti and G.Sardanashvily, {\it Gauge
Mechanics} (World Scientific, Singapore, 1998).

%15
\bibitem{dew} A.Dewismw and S.Bouquet, J. Math. Phys. {\bf
34}, 997 (1993).

%16
\bibitem{jmp00} L.Mangiarotti and G.Sardanashvily, J. Math. Phys. {\bf
41}, 2858 (2000).



%17
\bibitem{susm} H. Sussmannn, Trans. Amer. Math. Soc. {\bf 180},
171 (1973).



\end{thebibliography}
\end{document}